# *Benchmarking Vanilla GAN, DCGAN, and WGAN Architectures for MRI Reconstruction: A Quantitative Analysis*

Humaira Mehwish[1*], Hina Shakir[1], Muneeba Rashid[1], Asarim Aamir[1], Reema Qaiser Khan[2]
[1]Department of Software Engineering, Bahria University, Karachi, Pakistan;
hinashakir.bukc@bahria.edu.pk (H.M.) muneebarashid55@gmail.com (M.R.)
asarimsesodia15@gmail.com (A.A.).
[2]Department of Computer Science, Bahria University, Karachi, Pakistan;
reemaqaiser.bukc@bahria.edu.pk (R.Q.K.).
†These authors contributed equally to this work

*Corresponding author(s). E-mail(s): humairamehwishkhan@gmail.com;

## *Abstract*

*Magnetic Resonance Imaging (MRI) is a crucial imaging modality for viewing internal body structures. This research work analyses the performance of popular GAN models for accurate and precise MRI reconstruction by enhancing image quality and improving diagnostic accuracy. Three GAN architectures considered in this study are Vanilla GAN, Deep Convolutional GAN (DCGAN), and Wasserstein GAN (WGAN). They were trained and evaluated using knee, brain, and cardiac MRI datasets to assess their generalizability across body regions. While the Vanilla GAN operates on the fundamentals of the adversarial network setup, DCGAN advances image synthesis by securing the convolutional layers, giving a superior appearance to the prevalent spatial features. Training instability is resolved in WGAN through the Wasserstein distance to minimize an unstable regime, therefore, ensuring stable convergence and high-quality images. The GAN models were trained and tested using 1000 MR images of an anonymized knee, 805 images of Heart, 90 images of Brain MRI dataset. The Structural Similarity Index (SSIM) for Vanilla GAN is 0.84, DCGAN is 0.97, and WGAN is 0.99. The Peak Signal to Noise Ratio (PSNR) for Vanilla GAN is 26, DCGAN is 49.3, and WGAN is 43.5. The results were further statistically validated. This study shows that DCGAN and WGAN-based frameworks are promising in MR image reconstruction because of good image quality and superior accuracy. With the first cross-organ benchmark of baseline GANs under a common preprocessing pipeline, this work provides a reproducible benchmark for future hybrid GANs and clinical MRI applications.*

***Keywords:*** *Knee MRI, Heart MRI, Brian MRI, GANs, Vanilla GAN, DCGAN, WGAN, Image Reconstruction*

## *1. Introduction*

*Magnetic Resonance Imaging (MRI) is a non-invasive imaging technique widely used in medical diagnostics for obtaining high-resolution images of internal body structures, particularly soft tissues [1]. Unlike X-rays and CT scans that use ionizing radiation, MRI employs strong magnetic fields and*

*radio waves are used to generate detailed anatomical images, making them a safer and more effective choice for diagnosing soft tissue abnormalities.*

*While many knee disorders are treatable, accurate diagnosis is critical before initiating appropriate treatment. Similarly, in heart, arterial diseases, heart attacks, valve disorders, and even tumors and clots are important to detect precisely on time. Similarly for brain, tumors and injuries are crucial to diagnose. The size of the tumor, damage in soft tissues require a highly*

*accurate diagnosis. MRI plays a crucial role in diagnosing knee, brain and heart-related issues by capturing high-resolution images that provide detailed insights into the condition of bones and soft tissues.*

*However, one of the key limitations of MRI is the long acquisition time, which can lead to motion artefacts, increased cost, and patient discomfort. To mitigate these issues, researchers have explored various image reconstruction techniques that avoid rescanning and preserve image quality. The Generative Adversarial Networks (GANs) have gained attention for their ability to reconstruct high-quality images from under-sampled data, offering a promising solution to MR images.*

### 1.1. Challenges in Conventional MRI Reconstruction

*Despite its advantages, conventional MRI scanning suffers from several limitations. The process is time-consuming, often requiring patients to remain still for extended periods. Additionally, the high cost of MRI scans limits accessibility for many patients. The quality of MR images may also be affected by noise, motion artefacts, and low resolution, leading to potential inaccuracies in diagnosis [2].*

*The reconstruction of high-quality MR images is essential for enhancing anatomical analysis, facilitating early disease detection, and improving treatment planning. However, traditional reconstruction methods, such as Compressed Sensing MRI (CS-MRI) and iterative reconstruction techniques, may fail to produce optimal results regarding clarity, contrast, and detail preservation.*

### 1.2. Role of GANs in MR Images Enhancement

*AI-based models, such as Generative Adversarial Networks (GANs), leverage neural networks to generate high-quality images from undersampled or low-resolution MR images data [3] [4]. By learning complex spatial features and patterns, GANs can significantly improve image resolution, reduce noise, and enhance overall diagnostic accuracy.*

*This study evaluates three prominent GAN architectures: Vanilla GAN, Deep Convolutional GAN (DCGAN), and Wasserstein GAN (WGAN) for MR images reconstruction. These models are assessed based on their ability to reconstruct high-quality MR images with improved structural similarity, peak signal-to-noise ratio (PSNR), and inception score (IS).*

### 1.3. Contribution of the Research

*This research paper presents the following contributions while performing the reconstruction of MR images using Generative Adversarial Network (GAN) architectures:*

1. *The study utilizes the capabilities of Vanilla GAN, DCGAN and WGAN to address challenges associated with low-resolution or incomplete MR images, aiming to enhance image quality and resolution.*

2. *This comparison analyses the performance of these GANs on low-resolution MR images, offering deeper comprehension onto how generative models behave when image quality is low.*

3. *The findings aim to provide insights into the effectiveness of GAN-based frameworks in medical imaging and their potential to revolutionize MR images reconstruction techniques.*

*Although GANs have been used in the past to reconstruct MR images, fewer studies have conducted a direct, side-by-side comparison of Vanilla GAN, DCGAN, and WGAN based on the same preprocessing pipeline and dataset. In this study, the work was accomplished with a fair*

*experimental baseline and by comparing each GAN's performance on the same low-resolution knee, heart and brain MR images. Our findings give a solid baseline to build future GAN modifications and illustrate the strengths and limitations of every model, given real-world limitations.*

## 2. Related Work

*Most of the previous works within this domain have mainly dealt with knee MR imaging, which is the main area of this study. Though some related work is available for brain and cardiac MR images [4]. The heart and brain datasets were used experimentally to ensure the proposed GAN models' generalizability, whereas the literature review is still centered on knee MR images studies.*

### 2.1. Traditional and Non-GAN Deep Learning Techniques

*Knoll et al. [5] demonstrated the performance of adaptive intelligence techniques to accelerate Magnetic Resonance imaging processes. MR Knee k-space images were reconstructed using learning-based techniques and compressed sensing principles. An experimental dataset from the 2019 fastMRI challenge operated by Facebook AI Research and NYU Langone Health served as the foundation for training and testing on knee MR images data. Judith Herrmann et al. [6] analyzed deep-learning (DL) imaging of the knee at 1.5 and 3 T for diagnostic capability as well as high image quality. An evaluation of image quality was conducted using a Likert scale (1–5, with 5 representing the best outcome). TSEDL (Turbo Spin Echo-Deep Learning) images were rated as excellent (median 5, IQR 4–5), which outperformed the overall image quality ratings of TSEs (Turbo Spin Echo).*

*Chen et al. [7] presented PC-RNN (Pyramid Convolutional-Recurrent Neural Network) as a novel deep-learning method to perform reconstructive image processing with multiple scales. The framework designs a PC-RNN model that learns the features across multiple scales, following the inverse problem solution of MR images reconstruction. Johnson [8] evaluated the DL-image reconstruction prospectively for accelerated knee MR images to provide accurate diagnostics for internal knee disorders. An evaluation of abnormality detection capabilities between conventional images and DL-generated images was established through interchangeability tests.*

*Kaniewska et al. [9] measured the diagnostic accuracy of standard radial k-space (PROPELLER) MR images sequencing with accelerated data collections enhanced by deep learning- based convolutional neural network reconstruction for knee joint assessments. Ni et al. [10] examined AI-assisted compressed sensing (ACS) technology suitability for knee magnetic resonance imaging (MRI) scanning processes for enhancement and optimization. ACS technology demonstrates clear promise to serve as an alternative to traditional CS tools in 3D-MRI knee diagnostics because it enables high-speed imaging while maintaining diagnostic precision. Terzis et al. [11] examined the capability of Compressed Sensing (CS) when integrated with an AI-based super-resolution reconstruction prototype built from a series of convolutional neural networks (CNN) for executing an entire five-minute 2D knee MR images protocol. The study utilised two resolution settings (standard and low resolution) to obtain raw data through two reconstruction methods consisting of conventional Compressed SENSE (CS) and a new CNN-based approach for denoising and higher resolution image output (CS-SuperRes).*

### 2.2. GAN-Based Reconstruction Approaches

*Ke Lei et al. [12] applied unpaired adversarial training in reconstruction networks by pairing undersampled k-space data and reconstructed high-quality images. The networks contain generators for artefact reduction and discriminators that use disparate labels to optimize reconstruction performance. The generator functions as an unrolled neural network by*

*combining convolutional layers and data consistency layers in succession. Through its design as a multilayer CNN structure, the discriminator functions together with the critic and uses the Wasserstein distance to generate quality scores for image reconstruction. Sandilya et al. [13] addressed the reconstruction of MR images by proposing a Generative Adversarial Network (GAN) technique. The proposed approach yielded better results than common CS-MRI combined with modern methods through both quantitative metrics PSNR and SSIM index, and BRISQUE and FID score evaluation, as well as qualitative metrics mean opinion score and LPIPS evaluation.*

*Ma, Y et al. [14] developed a data structure technique using U-Net and a conditional generative adversarial network (CGAN) for MR image-based knee model creation, which produced accurate local SAR (Specific Absorption Rate) calculations by constructing proper knee models. Malczewski et al. [15] presented a generative multilevel network that served as the proposed solution for training 3D neural networks that operate with deficient MR input data. The authors proposed super-resolution reconstruction (SRR) alongside modified sparse sampling. Image-based Wasserstein GANs retain k-space data sparsity.*

*Patel et al. [16] researched a combination of analytical super-resolution (SR) approaches designed to improve the resolution quality of medical magnetic resonance imaging (MRI) scans. These sequential operations form the basis of the pipeline. Vallejo-Cendrero et al. [17] introduced a CycleGAN model for MR-to-pseudo-CT synthesis for knee imaging. The aim was to produce realistic pseudo-CT images from knee MR images scans that can reduce the use of individual CT scans in musculoskeletal imaging pipelines.*

## 2.3. Hybrid or Transformer-Based Architectures

*Hongki Lim [18] proposed a transformer-based combined framework for accelerated knee MR images reconstruction and segmentation jointly. The model offered better image reconstruction and segmentation performance, giving it potential use in clinical applications for enhancing MR images diagnostic quality and efficiency.*

*Zhang et al. [19] proposed FPS Former, a Frequency Pyramid Transformer model for accelerated knee MR images reconstruction. The FPS Former combines frequency-domain feature extraction with transformer-based attention to better capture local and global anatomical features.*

## 2.4. Comparative Benchmarking for Standard GANs

*While the aforementioned studies present innovative methods, however, they typically focus on a single model and applications without offering a comparison across standard GAN variants. Our work systematically benchmarks Vanilla GAN, DCGAN, and WGAN on the same knee MR images dataset using consistent metrics. This helps to establish stability, supporting future development of advanced, sophisticated models.*

## 3. Generative Adversarial Network Architectures for MR images Reconstruction

*Generative Adversarial Networks (GANs) are a class of deep learning models that generate high-quality synthetic images by training a generator and a discriminator in a competitive framework. In medical imaging, GANs have been successfully applied to reconstruct undersampled MR images scans, enhance image resolution, and remove noise and artefacts. The ability of GANs to synthesize realistic images makes them highly suitable for improving MR image reconstruction quality.*

### 3.1 Vanilla GAN

*A Generative Adversarial Network (GAN) training involves two components: the generator and*

*discriminator [22, 23]. The architecture of Vanilla GAN is shown in Figure 1. It begins with a noise vector by running through a dense layer and then being reshaped. The convolution layers are transposed, increasing the spatial dimensions of the tensor, and the final output is an image. To enhance model training and balance the spatial and channel dimensions, we apply normalization after each convolution layer, and LeakyReLU activation to avoid dead neurons, which only allow small negative values. The target of the generator is to generate actual pictures from the noise.*

*The discriminator assesses the fakeness or realness of images (images that the generative model synthesizes). Convolutional layers down-sample the image while extracting spatial features, and the number of filters increases in the last layer of the block, accepting an image. LeakyReLU creates non-linearity, and the dropout layers prevent overfitting. To predict the probability of the feature map being activated, the feature map is reshaped to a 1D vector, followed by a dense layer with sigmoid activation. The main parameters of the model are trained using binary cross-entropy loss during training, while the model of the generator receives feedback from the discriminator. This strengthens the generator's capacity to produce strongly dissimilar images while the discriminator enhances the rapid identification of fake images.*

*The parameter $L_D$ is used to evaluate the discriminator's capacity to distinguish between real and manufactured samples.*

$$L_D = -\frac{1}{m}\sum_{i=1}^{m} \log D(x_i) - \frac{1}{m}\sum_{i=1}^{m} \log(1 - D(G(z_i))) \quad\quad Eq.(1)$$

*Here $(xi)$ is the log-likelihood that the discriminator will correctly classify real data.*
*Log $(1 - ((zi)))$ is the chance that the discriminator will properly classify generated data as false.*

*To minimise this loss, the discriminator's job is to distinguish between actual and fake samples precisely. The MinMax Loss formula in a Generative Adversarial Network (GAN) is given by:*

$$minG\ maxD\ (G, D) = \mathbb{E}x\sim[\log D(x)] + \mathbb{E}z\sim_{\rho_z(z)}[\log(1 - D(g(z)))] \quad\quad Eq.(2)$$

*Where the discriminator network is D, and the generator network is G. $x$ represents actual data samples that were taken from the true data distribution $\rho_{data}(x)$. Random noise taken from an earlier distribution sample $(z)$ is a representation of z, which is often a normal or uniform distribution. The discriminator's probability of accurately classifying actual data as real is represented by $(x)$. The probability that the discriminator will recognize generated data from the generator as authentic is $(z)$. $\mathbb{E}_{x\sim\rho data}$ is the expectation (average) over real data samples x drawn from the real data distribution*

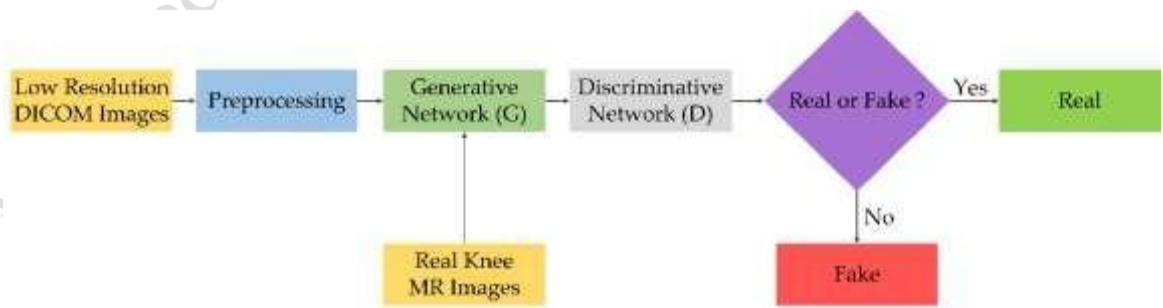

*Figure 1. Vanilla GAN Architecture (Generator and Discriminator)*

$\rho data$. $\mathbb{E}_{z\sim(z)}$ is the expectation (average) over random noise *z* drawn from distribution $\rho_z$. $log(1 − D(g(z)))$ is the log probability that the discriminator correctly identifies fake data as fake.

Vanilla GAN works to generate realistic medical images, but it also shows shortcomings in training fluctuation and mode dropping. It becomes a sound starting point for judging enhancements when it comes to other types of GANs in terms of image quality and consistency during training time.

### 3.2. DCGAN

The Deep Convolutional GAN (DCGAN) builds upon Vanilla GAN by replacing fully connected convolutional layers [22]. These convolutional layers allow the model to capture spatial dependencies more effectively. This is particularly useful for critical tasks such as MR images reconstruction, where preserving fine anatomical details is crucial. Additionally, batch normalization is integrated into DCGAN to stabilize training, making it more suitable for high-quality images generation.

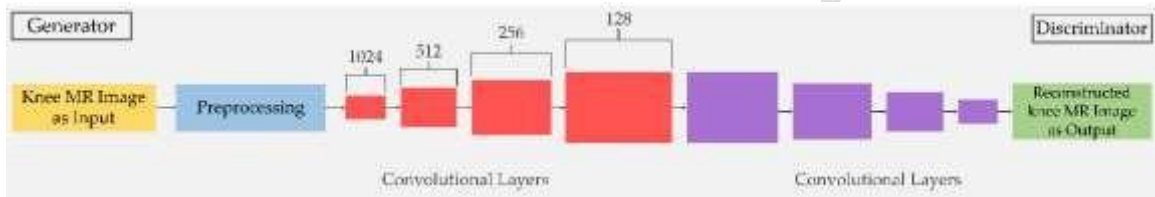

*Figure 2. Architecture of DC GAN*

These architectural improvements make DCGAN a strong candidate for assessing the impact of convolutional layers on the quality and stability of MR images reconstructions. Figure 2 shows the architecture of DCGAN.

The loss equation for DCGAN is identical to that of Vanilla GAN, shown as Eq. (2). Here is the output equation for the DCGAN given below:

$$[n] = h[n] * x[n] = \Sigma(h[k] * x[n − k]) \quad\quad Eq.(3)$$

Equation (3) illustrates the standard convolution operation that forms the basis of DCGAN's convolutional layers, aids in boosting spatial feature extraction for MR images reconstruction. Here, $y[n]$ is the output signal at time index n. $h[n]$ is the impulse response of the filter at time index n $x[n]$ is the input signal at time index n. $\Sigma$ denotes the summation over all possible values of k. At a particular time index, the output is equal to the weighted sum of input samples and their corresponding impulse response values. The coefficients of the impulse response give the weights for this weighted sum.

### 3.3. Wasserstein GAN (WGAN)

The Wasserstein Generative Adversarial Network (WGAN) improves the training stability of GANs by replacing the binary cross-entropy loss with Wasserstein loss [23]. This alternative loss function provides continuous feedback, enabling smoother gradient flow throughout the training process. As a result, WGAN enhances the model's convergence and mitigates issues such as mode collapse. This makes WGAN particularly effective for generating high-resolution images, including applications like MR images reconstruction, where stability and quality are crucial.

The discriminator needs modifications to eliminate the output layer's sigmoid function for producing scalable values suitable for Wasserstein loss computation. So, the role of the discriminator is replaced with a "critic," which estimates the Wasserstein distance rather than classifying data as real or fake. Weight clipping and gradient penalty employed by WGAN enforce Lipschitz continuity to maintain stable operation of the Wasserstein metric. A reconfigured training system updates the discriminator model more often than the generator model to improve final convergence results. The modifications to WGAN resolve common pathologies of Vanilla GANs, resulting in a robust and efficient alternative for generative modelling. The Wasserstein loss $W(\mathbb{P}_r, \mathbb{P}_g)$ can be computed using Eq.(4) as follows:

$$W(\mathbb{P}_r, \mathbb{P}_g) = \inf_{\gamma \in \Pi(\mathbb{P}_r, \mathbb{P}_g)} \mathbb{E}_{(x,y)\sim\gamma}[||x-y||] \qquad Eq.~(4)$$

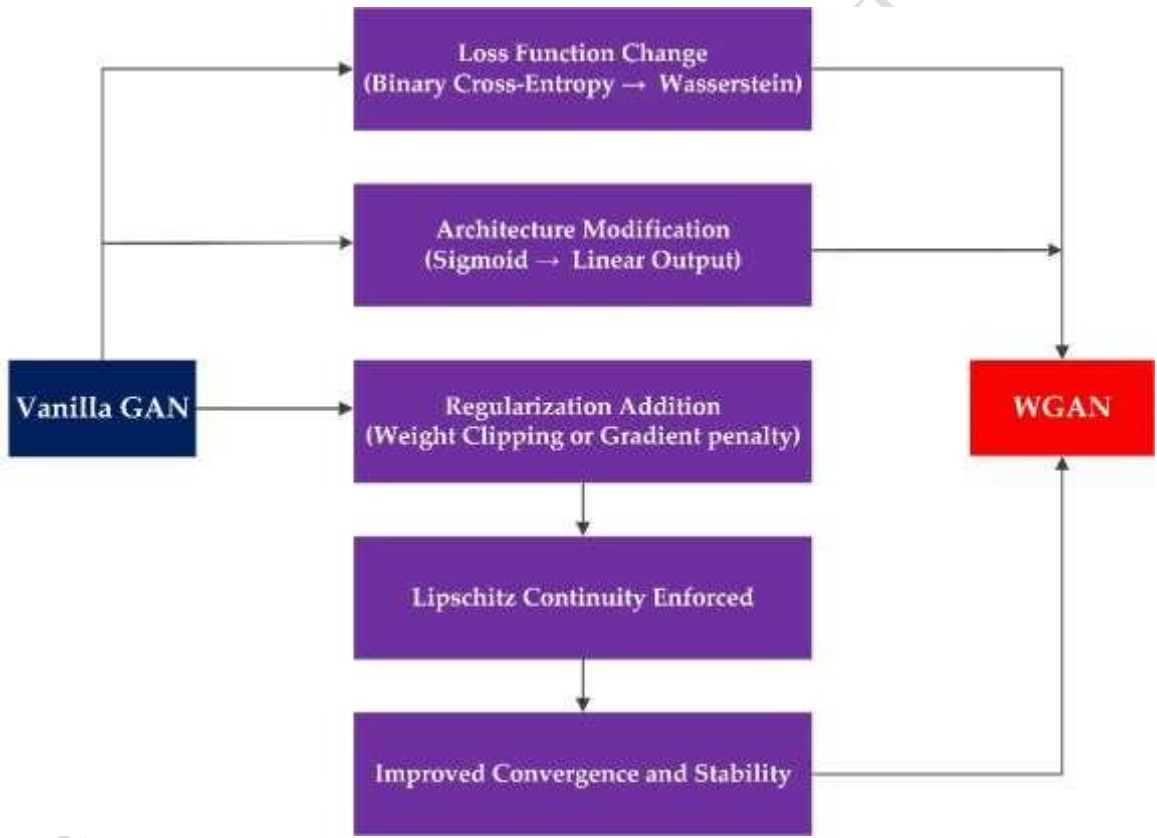

*Figure 3. Architecture of WGAN*

Where $\gamma$ represents the amount of mass moved from $x$ to $y$ to convert the distribution $P_r$ (real data distribution) into $P_g$ (generated data distribution). The collection of all possible joint distributions $\gamma(x,y)$, where the marginal distributions correspond to $P_r$ and $P_g$, respectively. WGAN's use of Wasserstein loss results in smoother and more reliable gradients, improving convergence during training. To satisfy the mathematical requirements of the Wasserstein distance, WGAN clips the critic's weights to enforce Lipschitz continuity, preventing instability.

## 4. Research Methodology

This study compares three GAN architectures: Vanilla GAN, DCGAN, and WGAN. Vanilla GAN

*uses a basic adversarial setup with fully connected layers, whereas DCGAN incorporates convolutional layers to enhance spatial feature extraction. WGAN further improves training stability by employing the Wasserstein loss function. Performance evaluation includes image quality, training stability, and convergence speed using metrics such as PSNR, IS, and SSIM.*

### 4.1. Experimental Dataset

*The experimental dataset comprises knee, heart and brain is reported in Table 1.*

#### a) Knee Dataset

*The knee and consists of 1000 anonymized knee MR images scans from the NYU Langone Health database, in DICOM format, containing raw k-space data [28]. This dataset is licensed for open research, provides k-space data that represents spatial frequency information of knee structures, including the femur, tibia, and patella. The k-space data is used for research on musculoskeletal conditions such as osteoarthritis and ligament injuries. The dataset is divided into 70% training and 30% testing subsets. Over-fitting was minimised by continuously monitoring training and test loss curves as well as evaluation measures (SSIM, PSNR) to ensure consistent convergence. 30% test set provides a reasonable basis for assessing model generalization. Future work may incorporate cross-validation or regularization techniques for further enhancement.*

#### b) Heart Dataset

*The Sunnybrook Cardiac Data (SCD) [29], also known as the 2009 Cardiac MR Left Ventricle Segmentation Challenge data, consists of 45 cine-MRI images from a mixed of patients and pathologies: healthy, hypertrophy, heart failure with infarction and heart failure without infarction. It consists of 805 MR images. A subset of this data set was first used in the automated myocardium segmentation challenge from short-axis MR images, held by a MICCAI workshop in 2009. The complete dataset is now available in the CAP database with a public domain license.*

#### c) Brain Dataset

*The Magnetic Resonance (MR) Jordan University Hospital (JUH) dataset [30] has been collected after receiving Institutional Review Board (IRB) approval of the hospital, and consent forms have been obtained from all patients. All procedures have been carried out in accordance with The Code of Ethics of the World Medical Association (Declaration of Helsinki).*

*The dataset consists of 2D image slices extracted using the RadiAnt DICOM viewer software. The extracted images are transformed to DICOM image data format with a resolution of 256x256 pixels. There are a total of 179 2D axial image slices referring to 20 patient volumes (90 MR). The dataset contains MR brain tumour images with corresponding segmentation masks. The MR images of each patient were acquired with a 5.00mm T Siemens Verio 3T using a T2-weighted sequence without contrast agent, 3 Fat sat pulses (FS), 2500-4000 TR, 20-30 TE, and 90/180 flip angle. The MR scans have a resolution of $0.7x0.6x5\ mm^3$.*

| *Data Source* | *Number of Images* | *Image Dimensions* | *Categories* | *Training/Test Split* |
|---|---|---|---|---|
| *NYU [26]* | *100* | *64x64* | *Knee MRI* | *70% - 30%* |
| *SCD [28]* | *805* | *256x256* | *Heart MRI* | *70% - 30%* |
| *JUH [27]* | *95* | *256x256* | *Brain MRI* | *70% - 30%* |

*Table 1: Description of Experimental Data Set*

## 4.2. Flow of Data through GANs

Figure 4 represents the flow of data through different types of Generative Adversarial Networks (GANs) in terms of their performance on a common image reconstruction task.

### 4.2.1. Preprocessing

The dataset undergoes preprocessing, including normalization that efficiently handles large-scale data and standardizes it. The restructuring organizes the data into a meaningful and accessible format. Next, the annotation enriches the dataset with labels to enhance its analysis. Finally, the images are converted from DICOM to PNG format for comparability.

### 4.2.2. The GAN architectures

Pre- processed data is then input into three GAN architectures

- **Vanilla GAN**: Uses fully connected layers for image generation.
- **DCGAN**: Incorporates convolutional layers to enhance image quality.
- **WGAN**: Uses a critic instead of a discriminator to improve stability and realism.

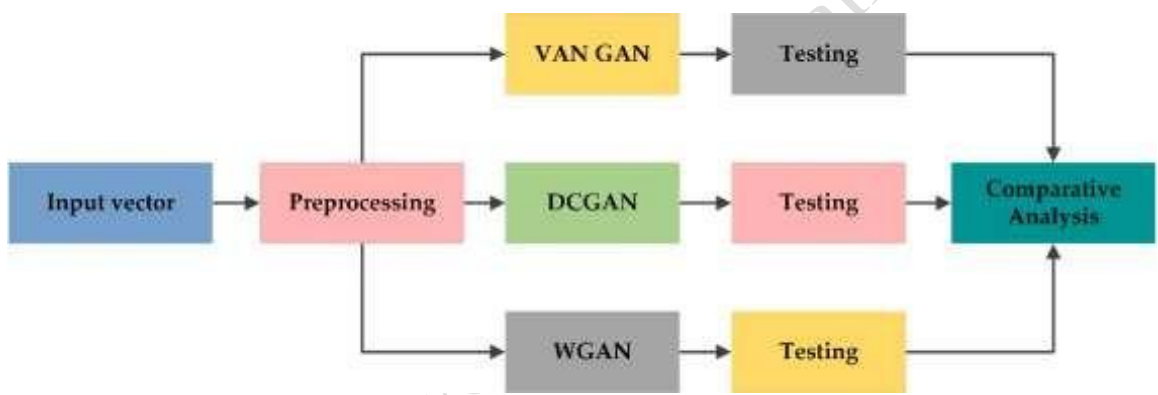

*Figure 4. Flow of research paradigm*

During testing, the generated images are evaluated using computational metrics such as SSIM and PSNR. The results from Vanilla GAN, DCGAN, and WGAN are compared to determine the most effective architecture.

## 4.3. Hyper-parameters Used in GAN Architectures

The methodology ensures a comprehensive comparison of GAN architectures for MR images reconstruction, providing valuable insights into their effectiveness in medical imaging. The hyper-parameters utilized in the GAN architectures are summarized in Table 1.

Vanilla GAN was trained for 2000 epochs, while DCGAN and WGAN were trained for 1000. Vanilla GAN exhibited slower and less stable convergence behavior during initial tests. The additional training epochs were used to enable its SSIM and PSNR metrics to stabilize. DCGAN and WGAN, on the other hand, converged faster with architectural enhancements by convolutional layers and Wasserstein loss.

|    | *Hyper-parameter* | *Vanilla GAN* | *DCGAN* | *WGAN* |
|----|-------------------|---------------|---------|--------|
| 1. | *Latent Dimension* | *100* | *100* | *100* |
| 2. | Batch Size | 128 | 128 | 128 |
| 3. | Number of Epochs | 2000 | 1000 | 1000 |
| 4. | Optimizer | Adam | Adam | Adam |
| 5. | Learning Rate | Generator: 0.00004 Discriminator: 0.0001 | Generator: 0.00005 Discriminator: 0.0002 | Generator: 0.00005 Discriminator: 0.0002 |

*Table 2: Hyper-parameters utilized in GAN Architectures*

## 5. Results and Comparative Analysis

*A visual comparison of three GAN based image reconstruction techniques for knee, heart and brain MR images is given in Figure 5, Figure 6 and Figure 7 respectively.*

### 5.1 Performance Metric
*The proposed GANs are compared based on the results obtained from the evaluation of the following research metric.*

#### 5.1.1 Structural Similarity Index (SSIM)

*SSIM is a metric used to assess the image quality of digital images. It's relatively more accurate in giving an impression related to human perception [23]. SSIM takes luminance, contrast, and structure into consideration. Luminance $I(x, y)$ measures the overall brightness, contrast $C(x, y)$ measures the difference between light and dark areas, and structure $S(x, y)$ evaluates how objects are organized within the image. Therefore, SSIM provides a meaningful assessment of image quality, especially when dealing with images that have been compressed or otherwise processed.*

$$I(x,y) = \frac{2\mu_x \mu_y + C_1}{\mu_x^2 + \mu_y^2 + C_1} \quad C(x,y) = \frac{2\sigma_x \sigma_y + C_2}{\sigma_x^2 + \sigma_y^2 + C_2} \quad S(x,y) = \frac{\sigma_{xy} + C_3}{\sigma_x \sigma_y + C_3} \quad Eq.\ (5)$$

$$(x, y) = (x, y)\ (x, y)(x, y)^\gamma \quad Eq.\ (6)$$

*The variables $\mu_x$, $\mu_y$, $\sigma_x$, and $\sigma_y$ denote the mean and standard deviations of pixel intensity in a local image patch centered at either x or y. Where $\alpha = \beta = \gamma = 1$*

*SSIM measures visual similarity between real and generated images across training epochs of Vanilla GAN, Deep Convolutional GAN (DCGAN), and Wasserstein GAN (WGAN). In Figure 8(a), the larger values represent higher perceptual quality. WGAN demonstrates the fastest and most uniform improvement, getting to almost perfect SSIM at epoch 1000 and then converging. DCGAN is very similarly trending, only slightly behind WGAN, but similarly achieving high stability in SSIM after 1000 epochs. On the contrary, Vanilla GAN also shows a far slower and more gradual recovery over epochs, after 2000 epochs. This reflects a much weaker capacity for maintaining structural image similarity compared to the other two models.*

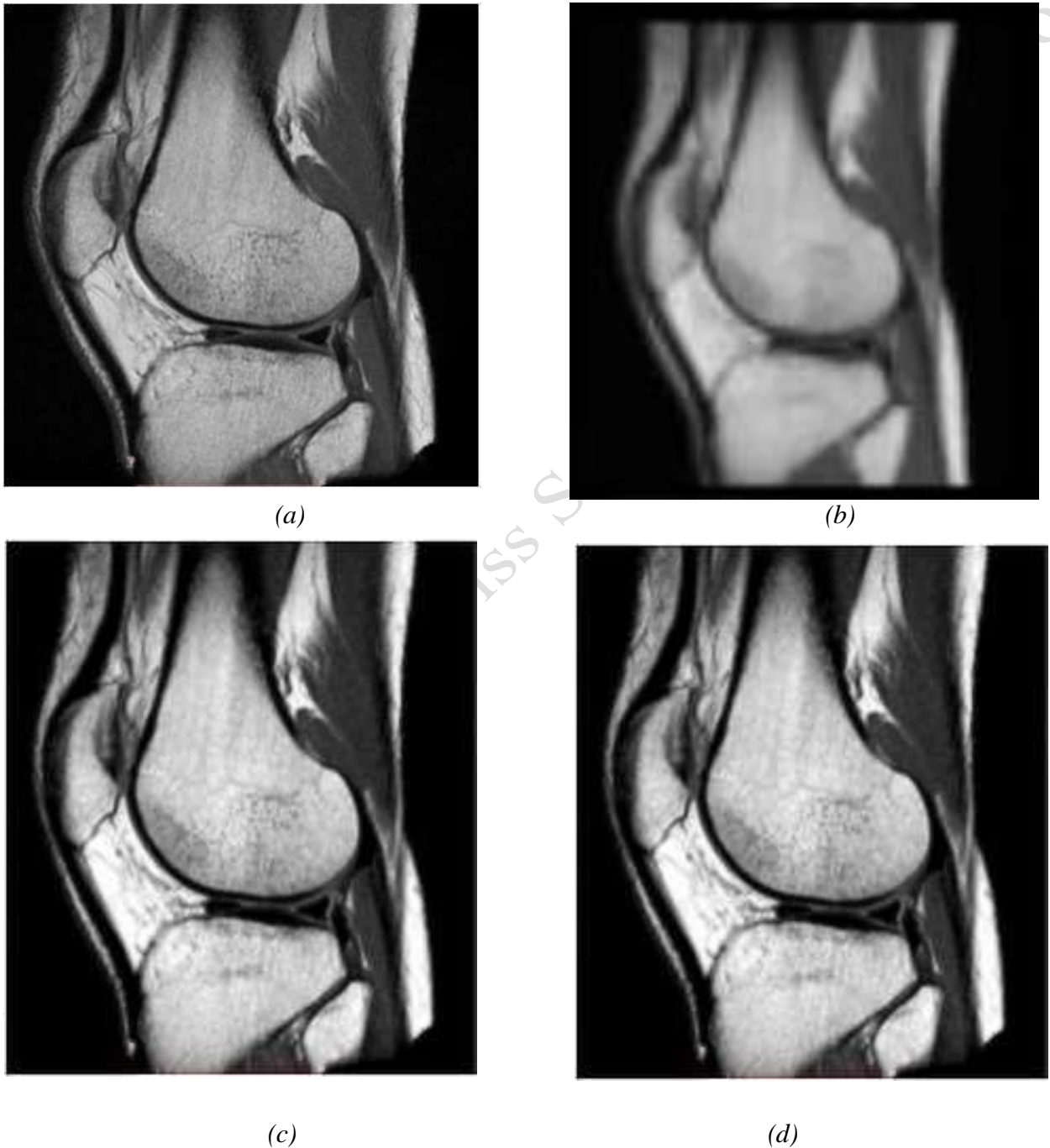

*(a)* *(b)*
*(c)* *(d)*

***Figure 5: Close-up visual comparison of the same knee slice reconstructed: (a) Original (b) Vanilla GAN reconstruction (c) DCGAN reconstruction (d) WGAN reconstruction***

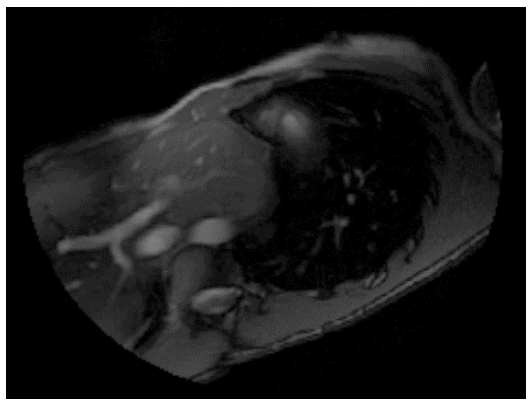
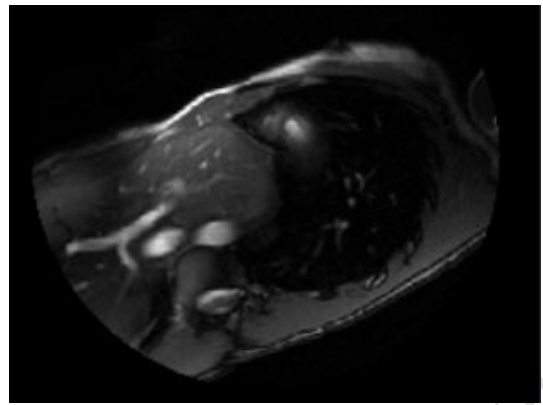

*(a)*            *(b)*

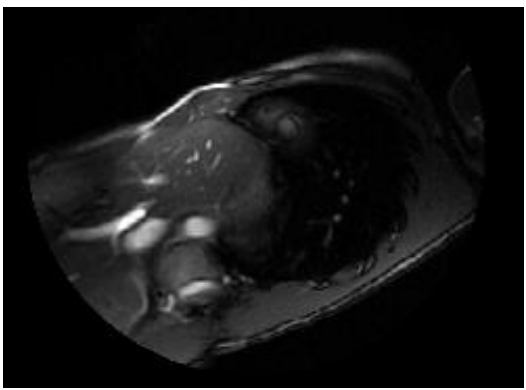
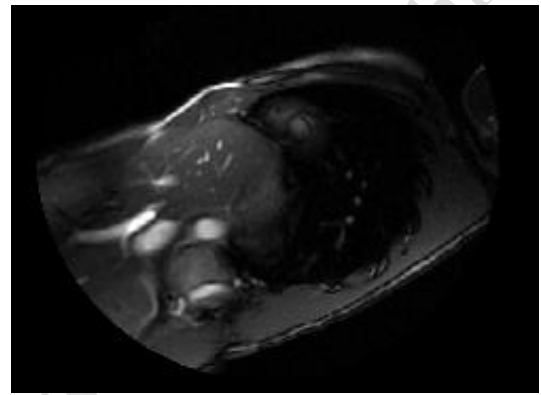

*(c)*            *(d)*

*Figure 6: Close-up visual comparison of the same Heart slice reconstructed: (a) Original (b) Vanilla GAN reconstruction (c) DCGAN reconstruction (d) WGAN reconstruction*

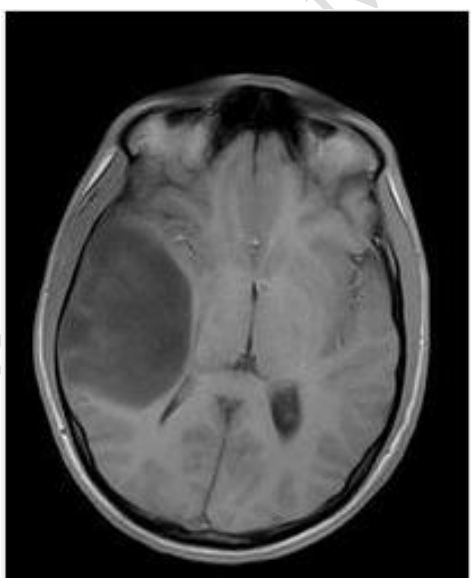
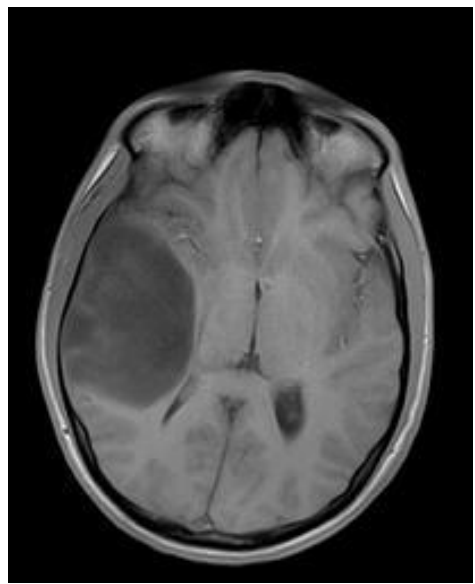

*(a)*            *(b)*

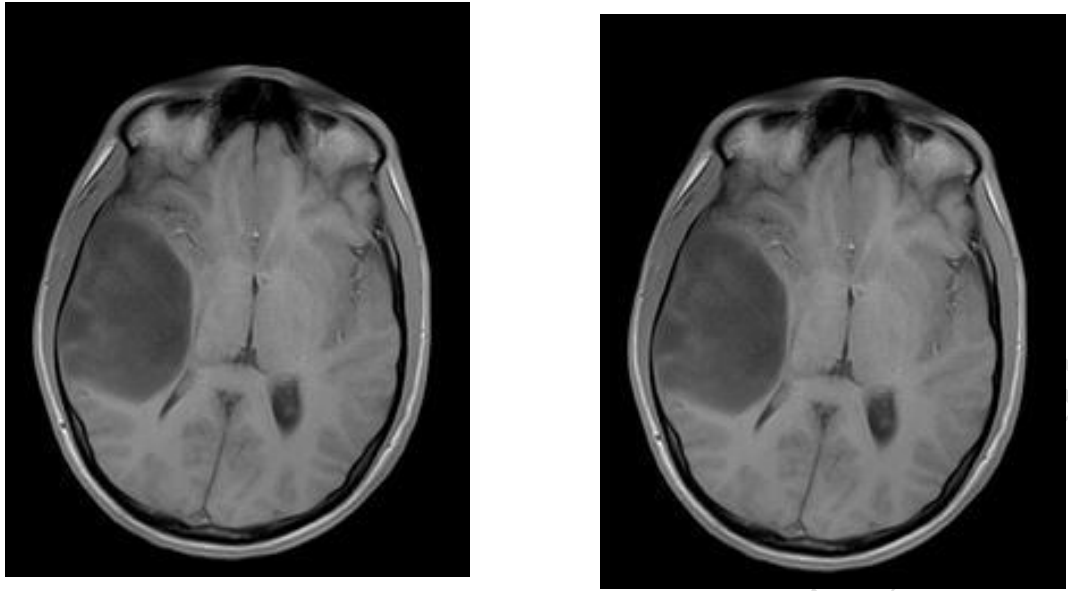

*(c)* *(d)*

*Figure 7: Close-up visual comparison of the same Brain slice reconstructed: (a) Original (b) Vanilla GAN reconstruction (c) DCGAN reconstruction (d) WGAN reconstruction*

### 5.1.2 *Peak Signal-to-Noise Ratio (PSNR)*

*Peak Signal-to-Noise Ratio (PSNR) measures the quality of reconstructed images against their originals. PSNR is computed from a comparison on a pixel-wise basis of the original and reconstructed images, taking the negative logarithm of Mean Squared Error (MSE). The better the quality of the reconstructed image, the higher the values of PSNR [23].*

$$PSNR(I, K) = 10 \log_{10}\left(\frac{MAX_I^2}{MSE}\right) \quad \text{Eq. (7)}$$

*Figure 8(b) displays a comparison of the performance of Vanilla GAN, DCGAN, and WGAN based on Peak Signal-to-Noise Ratio (PSNR) over training iterations. Both WGAN and DCGAN exhibit a steep rise in PSNR at the beginning of the training process, achieving an almost optimal value and sustaining that plateau, demonstrating fast convergence and high-quality image reconstruction. DCGAN improves PSNR by using convolutional layers for better feature extraction. Vanilla GAN exhibits a gradual and linear rise in PSNR over the epochs. However, WGAN achieves the highest PSNR by stabilizing training with Wasserstein loss, reducing artefacts. Overall, WGAN is preferred for high-quality image generation tasks.*

### 5.1.3 *Inception Score (IS)*

*In addition to employing conventional quantitative evaluation metrics, such as Structural Similarity Index (SSIM) and Peak Signal-to-Noise Ratio (PSNR), this study incorporates the Inception score (IS) [23] as an exploratory perceptual metric. The Inception Score assesses the quality of generated images in terms of their diversity and sharpness. While IS is predominantly utilized in the assessment of generated natural images to quantify visual realism and diversity, its application here serves as a supplementary measure aimed at evaluating the perceptual distinctiveness and visual fidelity of reconstructed MR images. According to Treder et al. [26], Inception Score (IS) has been shown to effectively assess both the quality and diversity of images generated by GANs, making it a valuable*

*tool in image quality evaluation. However, we acknowledge that IS is not specifically designed for grayscale medical images, and future studies may benefit from more domain-specific alternatives such as FID or clinical scoring.*

$$IS(G) = exp\ (\mathbb{E}_{x \sim p_y}[D_{\mathbb{KL}}\ (p(y|x)\ ||\ p(y))]) \qquad Eq.\ (8)$$

*where G is the generator model of the GAN. $x \sim p_y$ are generated images sampled from the generator's output distribution. $(y|x)$ is the conditional label distribution for image $x$ predicted by a pre-trained. Inception model, and $p(y)$ is the marginal distribution, calculated as the average of $(y|x)$ over many samples. $D_{\mathbb{KL}}$ is the Kullback-Leibler divergence, a measure of how one probability distribution diverges from another. Exp is the exponential function to transform the final score into a positive number.*

*Figure 8(c) gives a comparative evaluation of three Generative Adversarial Network (GAN) models: Vanilla GAN, DCGAN, and WGAN based on their Inception Score (IS) performance over training epochs. The outcomes reflect that both DC GAN and WGAN produce much higher IS values at a faster rate compared to Vanilla GAN, which suggests improved quality and diversity of the generated samples. DCGAN and WGAN both reach high IS values after the initial 1000 epochs and then stabilize, proving their stability and effectiveness in training. Vanilla GAN, on the other hand, increases slowly but steadily, never reaching even half the final score of the other two models, proving its inability to generate high-quality outputs. In general, DC GAN and WGAN perform much better than Vanilla GAN, and WGAN converges slightly better than DC GAN.*

### 5.1.4 Comparison of Performance Metric

*Table 3 shows the performance of Vanilla GAN, Deep Convolutional GAN (DCGAN), and Wasserstein GAN (WGAN), which is evaluated using Structural Similarity Index (SSIM), Inception Score (IS), and Peak Signal-to-Noise Ratio (PSNR). Vanilla GAN shows the lowest performance, with an SSIM of 0.84, IS of 2.9, and PSNR of 26, indicating poor structural similarity, lower image diversity, and higher noise. DCGAN significantly improves performance, achieving SSIM of 0.97, IS of 9.0, and PSNR of 43.5, due to its convolutional layers that enhance feature extraction and stability. WGAN matches DCGAN's performance, with SSIM of 0.99, IS of 9.0, and PSNR of 49.3, benefiting from the Wasserstein loss function that stabilises training and reduces mode collapse. Overall, both DCGAN and WGAN outperform Vanilla GAN, making them more suitable for generating high-quality images of three types of MR images (Knee, Heart, and Brain).*

### 5.2 Statistical measure -ANOVA Test

*One-way ANOVA [27] with post hoc tests were performed on Vanilla GAN, DC GAN and WGAN as these were independent groups. It was found that the f-value was 3079164.38 for SSIM and 18519133.33 for PSNR. These exceed the critical f-value for a 99.9% confidence for 200 observations and 3 groups, which suggests that these reconstruction techniques substantially affect the quality of the image enhancement, with some techniques outperforming others in terms of PSNR and SSIM. The results are significant at $p<0.1$. A post hoc Tukey's HSD statistic for each group pair was performed and shown in Table 2. This analysis confirms strong significant performance disparities in both PSNR and SSIM when comparing the Vanilla GAN technique against DC-GAN and WGAN. The WGAN result has demonstrated consistent superior image enhancement capabilities that are statistically significant.*

| | | *Post hoc Tukey's HSD for SSIM* | | | *Post hoc Tukey's HSD for PSNR* | | |
|---|---|---|---|---|---|---|---|
| *Group 1* | *Group 2* | *Mean Diff* | *Q-value* | *Significant* | *Mean Diff* | *Q-value* | *Significant* |
| Vanilla GAN | DCGAN | 0.12 | 2790.36(p<0.00001) | Yes | 17.32 | 139.48(p<0.00001) | Yes |
| Vanilla GAN | WGAN | 0.14 | 3238.97(p<0.00001) | Yes | 23.16 | 603.12(p<0.00001) | Yes |
| DCGAN | WGAN | 0.02 | 448.61 (p<0.00001) | Statistically Yes | 5.84 | 463.64 (p<0.00001) | Statistically Yes |

*Table 2: Calculating the statistical significance of quantitative measures significance*

### 5.3 Comparison with Other AI-Based MR images Reconstruction Techniques

*This comparison depicts a clearer insight into the performance of GAN models used in this research against a diverse set of MR images reconstruction techniques. Table 4 presents a comparative analysis of different AI techniques for MR images reconstruction, evaluating them based on Structural Similarity Index (SSIM), Inception Score (IS), and Peak Signal-to-Noise Ratio (PSNR). The presented Vanilla GAN achieves SSIM of 0.7, IS of 6.5, and PSNR of 26, showing lower performance compared to other methods. However, the designed DCGAN and WGAN outperform many existing techniques, achieving SSIM of 0.98, IS of 9.5, and PSNR of 42.5, demonstrating high image quality and structural similarity.*

*Compared to previous studies, GAN with Rician De-noising by Sandilya et al. [7] and U-Net and Conditional GAN by Ma et al. [8] have moderate performance, while Deep Learning Reconstruction (DLR) from Herrmann et al. [9] and Pyramid Convolutional RNN from Chen et al. 10] performed well but still fell short of the performance of the presented DCGAN and WGAN. Transformer-Based Integrated Framework (Hongki Lim, 2023)[14], Multilevel Generative Super-Resolution (Malczewski 2024)[15], and Image Domain Super-Resolution (Patel et al. 2024) [17] achieve competitive results but do not surpass the proposed WGAN and DCGAN in terms of SSIM and IS. These results highlight the effectiveness of the fine-tuned GAN-based methods in MR images reconstruction, ensuring better image fidelity and reconstruction quality.*

| GANs | SSIM | PSNR | IS |
|---|---|---|---|
| Vanilla GAN | 0.84±0.009 | 26±0.1 | 2.9 |
| DC GAN | 0.97±0.002 | 43±0.5 | 9.0 |
| WGAN | 0.99±0.002 | 49±0.3 | 9.0 |

*Table 3: Comparative Analysis using GAN Metrics*

'

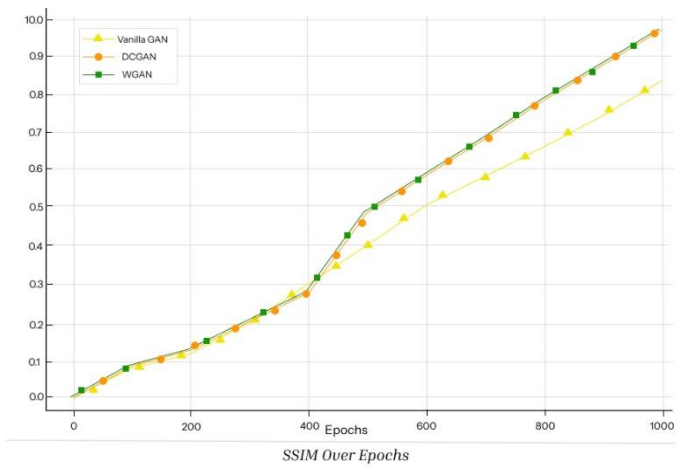

*(a)*

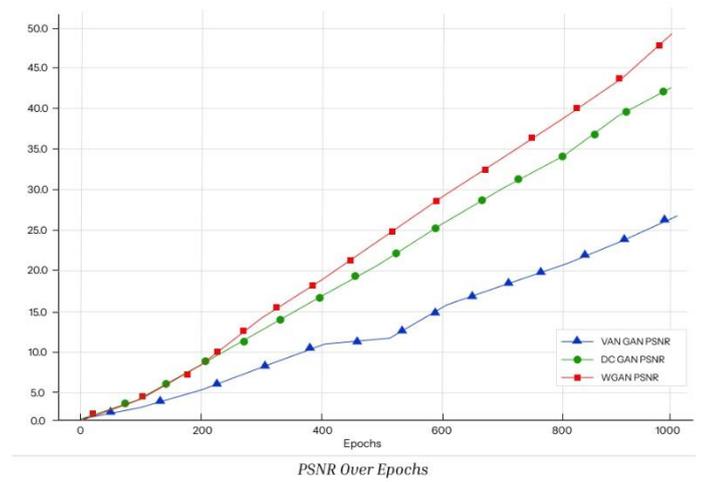

*(b)*

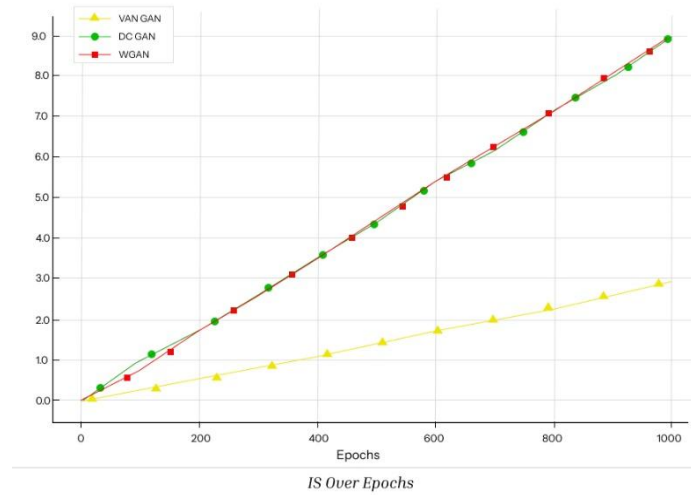

*(c)*

*Figure 8: Performance Comparison of Vanilla GAN, DCGAN and WGAN using: (a) SSIM over Epochs, (b) PSNR over Epochs, (c) IS over Epochs*

| Study | Technique | SSIM | PSNR |
|---|---|---|---|
| *Sandilya et al. (2021)[13]* | GAN with Rician De-noising | 0.89 | 32.10 |
| *Ma et al. (2021)[14]* | U-Net and Conditional GAN | 0.901 | 33.55 |
| *Herrmann et al.(2022)[6]* | Deep Learning Reconstruction (DLR) | 0.94 | 38.22 |
| *Chen et al. (2022)[7]* | Pyramid Convolutional RNN<br>Knee Single Coil<br>Knee Multi Coil | 0.72<br>0.92 | 32.35<br>38.75 |
| *Hongki Lim (2023)[18]* | Transformer-Based Integrated Framework | 0.83 | 35.55 |
| *Malczewski (2024)[15]* | Multilevel Generative Super-Resolution | Not Specified | 34.02 |
| *Zhang et al. (2024)[19]* | *FPS-Former (Frequency Pyramid Transformer) for fastMRI knee* | *0.90* | *32.85* |
| *Patel et al. (2024) [16]* | *Image Domain Super-Resolution* | *0.92* | *34.06* |
| *Vallejo-Cendrero et al.[17]* | *CycleGAN* | *0.92 ± 0.02* | *28.12 ± 1.52* |
| *Fine-tuned Vanilla GAN For (Knee, Heart, Brain)* | *Vanilla GAN* | *0.84±0.009* | *26±0.1* |
| *Fine-tuned DCGAN For (Knee, Heart, Brain)* | *DCGAN* | *0.97±0.002* | *43±0.5* |
| *Fine-tuned WGAN For (Knee, Heart, Brain)* | *WGAN* | *0.99±0.002* | *49±0.3* |

*Table 4: Comparative Analysis with other AI techniques for MR images reconstruction*

## 6. Conclusion

*The comparative analysis of Vanilla GAN, DCGAN, and WGAN for MR images reconstruction highlights key advantages and challenges associated with each model. Vanilla GAN serves as a foundational approach but struggles with training instability and lower image quality. DCGAN demonstrates significant improvements in accuracy and image clarity due to its use of convolutional layers, making it a strong candidate for medical image reconstruction. WGAN further enhances stability and reliability by leveraging Wasserstein loss, addressing issues like mode collapse and ensuring high-quality outputs.*

*One of the main novelties of this paper is to take benchmarking from knee MR images and extend it to brain and cardiac MR images datasets, thus showing the generalizability of GAN-based reconstruction from one anatomical site to others. This cross-organ comparison offers researchers and clinicians a broader baseline against which future work can be compared.*

## 7. Future Work

*Future studies must research more sophisticated variants like WGAN-GP, CycleGAN, and Attention-GAN, and also include radiologist ratings and clinical validation. Looking forward, these three supplementary variants can be compared to enhance performance and generalization. WGAN-GP will first substitute weight clipping with a gradient-penalty loss to regularize critic training and eliminate artefacts, and should achieve modest SSIM and PSNR improvements. CycleGAN will secondly allow to use unpaired datasets and counter domain shift using cycle-consistency, which is particularly useful where perfectly paired ground truth is not available. Third, Attention-GAN will add spatial and channel attention to the generator and discriminator so that the model attends to clinically relevant anatomy and maintains fine boundaries. Cumulatively, these directions should enhance stability, leverage larger unpaired cohorts, and increase detail fidelity without trading off anatomical realism.*

## Ethics declarations

### Ethics approval and consent to participate
*Not applicable. The MR images used in this study were obtained from the NYU Langone fastMRI database, which is a publicly available, anonymised dataset licensed for open academic research. As the data contains no personal identifiers and is ethically cleared by the original data providers, no*

*further ethical approval was required for this work.*

*Consent for publication*
*Not applicable.*

*Competing Interests*
*The authors declare that they have no financial or non-financial competing interests directly or indirectly related to the work submitted for publication.*

*Funding*
*This research received no specific grant from any funding agency, commercial entity, or not-for-profit organisation.*

*Code Availability*
*The trained models and code will be made accessible on reasonable request*
*for academic research purposes.*

*References*
[1] *Magnetic Resonance Imaging (MRI) (no date). https://www.nibib.nih.gov/science-education/science-topics/magnetic-resonance-imaging-mri.*
[2] *Foti, G. and Longo, C. (2024) 'Deep learning and AI in reducing magnetic resonance imaging scanning time: advantages and pitfalls in clinical practice,' Polish Journal of Radiology, 89, pp. 443–451. https://doi.org/10.5114/pjr/192822.*
[3] *Wolterink, J.M. et al. (2021) 'Generative Adversarial Networks: a primer for radiologists,' Radiographics, 41(3), pp. 840–857. https://doi.org/10.1148/rg.2021200151.*
[4] *Ali, H. et al. (2022) 'The role of generative adversarial networks in brain MRI: a scoping review,' Insights Into Imaging, 13(1). https://doi.org/10.1186/s13244-022-01237-0.\*
[5] *Knoll, F., Zbontar, J., Sriram, A., Muckley, M. J., Bruno, M., Defazio, A., Parente, M., Geras, K. J., Katsnelson, J., Chandarana, H., Zhang, Z., Drozdzalv, M., Romero, A., Rabbat, M., Vincent, P., Pinkerton, J., Wang, D., Yakubova, N., Owens, E., . . . Lui, Y. W. (2020). FastMRI: a publicly available raw K-Space and DICOM dataset of knee images for accelerated MR image reconstruction using machine learning. Radiology Artificial Intelligence, 2(1), e190007. https://doi.org/10.1148/ryai.2020190007*
[6] *Herrmann, J., Keller, G., Gassenmaier, S., Nickel, D., Koerzdoerfer, G., Mostapha, M., Almansour, H., Afat, S., & Othman, A. E. (2022). Feasibility of an accelerated 2D-multi-contrast knee MRI protocol using deep-learning image reconstruction: a prospective intraindividual comparison with a standard MRI protocol. European Radiology, 32(9), 6215–6229. https://doi.org/10.1007/s00330-022-08753-z*
[7] *Chen, E. Z., Wang, P., Chen, X., Chen, T., & Sun, S. (2022b). Pyramid convolutional RNN for MRI image reconstruction. IEEE Transactions on Medical Imaging, 41(8), 2033–2047. https://doi.org/10.1109/tmi.2022.3153849*
[8] *Johnson, P. M., Lin, D. J., Zbontar, J., Zitnick, C. L., Sriram, A., Muckley, M., Babb, J. S., Kline, M., Ciavarra, G., Alaia, E., Samim, M., Walter, W. R., Calderon, L., Pock, T., Sodickson, D. K., Recht, M. P., & Knoll, F. (2023). Deep learning reconstruction enables prospectively accelerated clinical knee MRI. Radiology, 307(2). https://doi.org/10.1148/radiol.220425*
[9] *Kaniewska, M., Deininger-Czermak, E., Lohezic, M., Ensle, F., & Guggenberger, R. (2023). Deep Learning Convolutional neural network reconstruction and Radial K-Space Acquisition*

[ ] *MR technique for enhanced detection of retropatellar cartilage lesions of the knee joint. Diagnostics, 13(14), 2438. https://doi.org/10.3390/diagnostics13142438*

[10] *Ni, M., He, M., Yang, Y., Wen, X., Zhao, Y., Gao, L., Yan, R., Xu, J., Zhang, Y., Chen, W., Jiang, C., Li, Y., Zhao, Q., Wu, P., Li, C., Qu, J., & Yuan, H. (2023). Application research of AI- AI-assisted compressed sensing technology in MRI scanning of the knee joint: 3D-MRI perspective. European Radiology, 34(5), 3046–3058. https://doi.org/10.1007/s00330-023-10368-x*

[11] *Terzis, R., Dratsch, T., Hahnfeldt, R., Basten, L., Rauen, P., Sonnabend, K., Weiss, K., Reimer, R., Maintz, D., Iuga, A., & Bratke, G. (2024). Five-minute knee MRI: An AI-based super resolution reconstruction approach for compressed sensing. A validation study on healthy volunteers. European Journal of Radiology, 175, 111418. https://doi.org/10.1016/j.ejrad.2024.111418*

[12] *Lei, K., Mardani, M., Pauly, J. M., & Vasanawala, S. S. (2020). Wasserstein GANs for MR Imaging: From paired to Unpaired training. IEEE Transactions on Medical Imaging, 40(1), 105–115. https://doi.org/10.1109/tmi.2020.3022968*

[13] *Sandilya, M., Nirmala, S. R., & Saikia, N. (2021). Compressed Sensing MRI Reconstruction Using Generative Adversarial Network with Rician De-noising. Applied Magnetic Resonance, 52(11), 1635–1656. https://doi.org/10.1007/s00723-021-01416-0*

[14] *Ma, Y., Xing, C., & Xiao, L. (2021). Knee Model Construction Based on MR images Using U-Net and Conditional GAN. IEEE. https://doi.org/10.1109/ICMIPE53131.2021.9698921*

[15] *Malczewski, K. (2024). A Framework for Reconstructing Super-Resolution Magnetic Resonance Images from Sparse Raw Data Using Multilevel Generative Methods. Applied Sciences, 14(4), 1351. https://doi.org/10.3390/app14041351*

[16] *Patel, V., Wang, A., Monk, A. P., & Schneider, M. T. (2024). Enhancing Knee MR Image Clarity through Image Domain Super-Resolution Reconstruction. Bioengineering, 11(2), 186. https://doi.org/10.3390/bioengineering11020186*

[17] *Vallejo-Cendrero, M.A., Pascau, J., Gálvez, M., Jiménez-Carretero, D., and García-Vázquez, V., 2024. MR-Based Pseudo-CT Synthesis for Knee Imaging Using CycleGAN. Applied Sciences, 14(11), p.4655. Available at: https://www.mdpi.com/2076-3417/14/11/4655*

[18] *Lim, H. (2023) 'Transformer-Based Integrated Framework for joint Reconstruction and Segmentation in Accelerated Knee MRI,' Electronics, 12(21), p. 4434. https://doi.org/10.3390/electronics12214434.*

[19] *Zhang, Y., Meng, Y., Yang, Z., and Shi, Y., 2024. FPS-Former: Frequency Pyramid Transformer for Fast MRI Reconstruction. arXiv preprint arXiv:2412.10776*

[20] *Yang, Y. (2021) 'Digit Image Generation with Vanilla GANs,' ResearchGate [Preprint]. https://www.researchgate.net/publication/356986216_Digit_Image_Generation_with_Vanilla_GANs.*

[21] *Generating images using vanilla generative adversarial networks (no date). https://ieeexplore.ieee.org/abstract/document/10840720*

[22] *DCGAN-based Pre-trained model for Image Reconstruction using ImageNet (no date). https://ieeexplore.ieee.org/abstract/document/9445128.*

[23] *Arjovsky, M., Chintala, S. and Bottou, L. (2017) Wasserstein GAN. https://arxiv.org/abs/1701.07875.*

[24] *Knoll et al Radiol Artif Intell . 2020 Jan 29;2(1):e190007. doi: 10.1148/ryai.2020190007. https://pubs.rsna.org/doi/10.1148/ryai.2020190007*

[25] *Borji, A. (2018) 'Pros and cons of GAN evaluation measures,' Computer Vision and Image Understanding, 179, pp. 41–65. https://doi.org/10.1016/j.cviu.2018.10.009.*

[26] *Treder, M.S., Codrai, R. and Tsvetanov, K.A. (2022) 'Quality assessment of anatomical MRI*

*images from generative adversarial networks: Human assessment and image quality metrics,' Journal of Neuroscience Methods, 374, p. 109579. https://doi.org/10.1016/j.jneumeth.2022.109579.*

[27] *Shakir, H., Deng, Y., Rasheed, H. et al. (2019) Radiomics based likelihood functions for cancer diagnosis. Sci Rep 9, 9501 https://doi.org/10.1038/s41598-019-45053-x*

[28] *FastMRI Dataset (no date). https://fastmri.med.nyu.edu/.*

[29] *Sunnybrook Cardiac MRI (2021). https://www.kaggle.com/datasets/salikhussaini49/sunnybrook-cardiac-mri.*

[30] *Unpaired MR-CT brain dataset (2024). https://www.kaggle.com/datasets/oalkadi/unpaired-mr-ct-brain-dataset.*